\documentclass{article}
\usepackage{amssymb}
\usepackage{amsmath}
\usepackage{chicago}

\newtheorem{theorem}{Theorem}
\newtheorem{acknowledgement}[theorem]{Acknowledgement}

\input{tcilatex}
\begin{document}

\title{Origin and meaning of quantum nonlocality}
\author{L. de la Pe\~{n}a,$^{\text{1}}$A. M. Cetto,$^{\text{1}}$ \and A. Vald%
\'{e}s-Hern\'{a}ndez$^{\text{1}}$, and H. M. Fran\c{c}a$^{\text{2}}$ \\
$^{1}$Instituto de F\'{\i}sica, \\
Universidad Nacional Aut\'{o}noma de M\'{e}xico, \\
A.P. 20-364, M\'{e}xico D.F., Mexico\\
e-mail: luis, ana, andreavh@fisica.unam.mx\\
$^{2}$Instituto de F\'{\i}sica, Universidade de S\~{a}o Paulo, \\
CP 66318, S\~{a}o Paulo, Brazil\\
e-mail: hfranca@if.usp.br}
\date{}
\maketitle

\begin{abstract}
Quantum nonlocality is revisited from a novel point of view by studying the
problem of an originally classical particle immersed in the stochastic
zero-point radiation field\ (zpf). The entire system is left to evolve until
it reaches a state in which the radiative terms cancel each other in the
mean in a first approximation. The ensuing approximate statistical
description reduced to the particle's configuration space contains a
nonclassical term due to the dispersion of the momentum, which depends on
the density of particles $\rho (x)$ and thus is nonlocal. This description
is shown to be equivalent to Schr\"{o}dinger's equation and its complex
conjugate. The nonlocal term is recognized as the so-called quantum
potential, thus solving the long standing problem of the origin and meaning
of this term. Further, the relationship between the Wigner function and a
true Kolmogorovian probability density in phase space is discussed from the
perspective provided by this theory.


\end{abstract}

\section{Introduction}

The aim of the present paper is to shed light on the physical origin and
meaning of quantum nonlocality, one of the major conceptual quandaries of
quantum mechanics. We accomplish this within a theoretical framework that
identifies the presence of the stochastic zero-point radiation field (zpf)
in interaction with matter as the entity that is ultimately responsible for
the fluctuating properties of the particle's momentum, which from this
perspective are at the root of both the nonlocal and the indeterministic
nature of the quantum description.

The present paper is based on previous work,$^{\text{\cite{DeFrMa98}-\cite%
{PeCe77}}}$ but supersedes it in a significant way. It starts by considering
the full (infinite dimensional), nonrelativistic and stochastic phase-space
statistical description of the field and particle in interaction. After
averaging over the realizations of the zpf the{\LARGE \ }system is allowed
to evolve towards a state of mean energy balance between the particle and
the zpf, which takes us in the radiationless limit to Schr\"{o}dinger's
theory, once the description is reduced to the configuration space of the
particle. Of particular importance is the observation that this reduction of
the description from the complete phase space to the configuration space of
the particle gives rise to a term representing the fluctuations of the
momentum impressed by the zpf on the particle. This term is a function of
the probability density $\rho (x)$ and hence possesses a nonlocal nature,
manifested even in one-particle quantum systems. The origin of the quantum
nonlocality appearing in the reduced description is thus traced to the
action of the fluctuating background field{\footnotesize .}\footnote{%
The term nonlocality is used here in its standard sense, to refer to a
direct, instantaneous (nonmediated) influence between two distant objects or
points in space. The zpf, as any Maxwellian field, obeys of course the
principle of locality.\ } Planck's constant $\hbar $ enters into the
quantum-mechanical description precisely through the nonlocal term related
to the momentum dispersion, as a measure of the fluctuations driven by the
zpf.

It is further shown that under the stated approximations the phase-space
probability density --- which is explicitly written --- is transformed into
the Wigner function. In the course of the approximations and partial
averaging the original true (that is, Kolmogorovian) probability density
loses some of its properties, which explains why the phase space probability
functions employed in the quantum description (and particularly the Wigner
function) are normally not true probability densities.

The results presented imply that the quantum phenomenon is not\ innate ---
it is not intrinsic to matter or to the radiation field alone --- but is an
emergent phenomenon generated by the permanent matter-field interaction.
This observation reinforces the results obtained in Refs. \cite{PeVaCe08}-%
\cite{VaPeCe11}. In the first two of these references it is shown that the
mere introduction of the zero-point energy of the radiation field\ into an
otherwise classical description suffices to explain the Planck distribution
and the discrete properties of the field. A detailed study of the dynamics
of the mechanical system immersed in the zpf is shown to lead to the
Heisenberg formalism of nonrelativistic quantum mechanics.$^{\text{\cite%
{PeCe07},\cite{PeVaCe09}}}$ and to entanglement of particles.$^{\text{\cite%
{PeVaCe10},\cite{VaPeCe11}}}${\LARGE \ }Altogether, these results show that
a careful study of (originally classical) material systems in interaction
with\ a stochastic radiation field endowed with a zero-point contribution
allows to achieve a better understanding of the quantum description. A short
version of the present paper is given in Ref. \cite{PeVaCeFr11}.

\section{Reducing the Liouville Equation}

Our starting point is the Liouville equation for the density $R$ of points $%
(x_{s},p_{s})$ in the phase space of the complete system composed of a
particle (of mass $m$\ and charge $e$) and the zero-point field (here $x_{s}$
and $p_{s}$ stand for the whole set of degrees of freedom $x_{s}=\left\{ x_{%
\text{part}}\equiv x,x_{\text{zpf}}\right\} ,$ $p_{s}=\left\{ p_{\text{part}%
}\equiv p,\text{ }p_{\text{zpf}}\right\} $): 
\begin{equation}
\frac{\partial }{\partial t}R(x_{s},p_{s},t)+\frac{\partial }{\partial x_{s}}%
\left( \overset{\cdot }{x_{s}}R(x_{s},p_{s},t)\right) +\frac{\partial }{%
\partial p_{s}}\left( \overset{\cdot }{p_{s}}R(x_{s},p_{s},t)\right) =0.
\label{30a}
\end{equation}%
The motion of the mechanical (one-dimensional) subsystem is governed, in the
nonrelativistic limit, by 
\begin{equation}
\dot{x}=p/m,\text{ \quad }\dot{p}=f(x)+m\tau \dddot{x}+eE(t),  \label{30b}
\end{equation}%
where $f(x)$ is the external force, $E(t)=\mathbf{E}(t)\cdot \mathbf{\hat{x}}
$ with $\mathbf{E}(t)$ the electric component of the zpf in the
long-wavelength approximation and $m\tau \overset{...}{x}$ is the radiation
reaction force, with $\tau =2e^{2}/3mc^{3}$ ($\approx 10^{-23}$ s for the
electron). We will further approximate this term in the form (although this
will be of little importance in what follows) $m\tau \dddot{x}\simeq \tau 
\dot{x}f^{\prime }(x),$ where the prime denotes derivative with respect to $%
x $.

We are not interested in the motions for a certain realization of the random
field $E(t)$, but in the average motion of an ensemble of similar systems.
We therefore average $R(x_{s},p_{s},t)$ over the realizations of this field
and call $Q(x,p,t)$ the averaged density in the phase space of the particle.
The procedure can be performed using the projector technique (see e.g. Refs 
\cite{PeCe77} and \cite{Frisch68}). After a somewhat lengthy calculation
Eqs. (\ref{30a}) and (\ref{30b}){\LARGE \ }lead to the Fokker-Planck-type
equation 
\begin{equation}
\frac{\partial }{\partial t}Q+\frac{1}{m}\frac{\partial }{\partial x}pQ+%
\frac{\partial }{\partial p}\left[ \left( f(x)+\frac{\tau }{m}f^{\prime
}(x)p\right) Q\right] =e^{2}\frac{\partial }{\partial p}\hat{D}(t)Q,
\label{32}
\end{equation}%
where $\hat{D}(t)$ is a complicated integro-differential operator. More
precisely the `diffusion' term can be written in the form$^{\text{\cite%
{PeCe77}}}$%
\begin{equation}
e^{2}\frac{\partial }{\partial p}\hat{D}(t)Q=e\frac{\partial }{\partial p}%
\hat{P}E\sum_{k=0}^{\infty }\left[ e\hat{G}\frac{\partial }{\partial p}(1-%
\hat{P})E\right] ^{2k+1}Q.  \label{32b}
\end{equation}%
$\hat{P}$ stands for the smoothing operator that averages over the
realizations of the field ($\hat{P}A=\overline{A}^{E}$) and $\hat{G}$
represents the operator 
\begin{equation}
\hat{G}A(x,p,t)=\left( \frac{\partial }{\partial t}+\hat{L}\right)
^{-1}A(x,p.t)=\int_{0}^{t}e^{-\hat{L}(t-t^{\prime })}A(x,p,t^{\prime
})dt^{\prime },  \label{32c}
\end{equation}%
($\hat{L}=\frac{1}{m}\frac{\partial }{\partial x}p+\frac{\partial }{\partial
p}(f+\tau f^{\prime }\frac{p}{m})$). The explicit form of $\hat{D}(t)$ will
not be required for the present work. However, it is important to note that
the diffusion term (\ref{32b}) comprises at least a correlation of the form $%
E(t)E(t^{\prime })$\ averaged over the realizations of the field, and hence
contains a factor proportional to the spectral (energy) density of the zpf,
given by{\LARGE \ }

\begin{equation}
\rho _{\text{zpf}}(\omega )=\frac{\hbar \omega ^{3}}{2\pi ^{2}c^{3}}=\frac{%
\omega ^{2}}{\pi ^{2}c^{3}}\cdot \frac{\hbar \omega }{2},  \label{34}
\end{equation}%
as follows from the fact that the energy per normal mode of frequency $%
\omega $ is $\hbar \omega /2.$ (Incidentally, we recall that this expression
is derived without quantum assumptions from the demand of Lorentz
invariance; see Ref. \cite{dice} and references therein).

\section{Transition to configuration space}

Our interest here lies in the connection of Eq. (\ref{32}) with quantum
mechanics, which is usually described in configuration space.\footnote{%
Here we restrict ourselves to the description in configuration space. A
parallel process can of course be followed to reduce the description to
momentum space, performing the Fourier transformation in configuration space
instead.} The transition to configuration space can be performed
systematically by means of the characteristic (moment generating) function $%
\widetilde{Q}$ associated with the density $Q,$%
\begin{equation}
\widetilde{Q}(x,z,t)=\int Q(x,p,t)e^{ipz}dp.  \label{40}
\end{equation}%
\thinspace From this expression it follows that the probability density $%
\rho (x,t)$, or rather the marginal probability in configuration space, is 
{\LARGE \ } 
\begin{subequations}
\begin{equation}
\rho (x,t)=\int Q(x,p,t)dp=\widetilde{Q}(x,0,t),  \label{42a}
\end{equation}%
and the (partially averaged) local moments of $p$ are given by 
\begin{equation}
\left\langle p^{n}\right\rangle (x)\equiv \left\langle p^{n}\right\rangle
_{x}=\frac{1}{\rho (x)}\int p^{n}Qdp=(-i)^{n}\left. \left( \frac{1}{%
\widetilde{Q}}\frac{\partial ^{n}\widetilde{Q}}{\partial z^{n}}\right)
\right\vert _{z=0}.  \label{42b}
\end{equation}%
Now we introduce the change of variables 
\end{subequations}
\begin{equation}
z_{+}=x+\beta z,\text{ \quad }z_{-}=x-\beta z,  \label{52}
\end{equation}%
with an as yet undetermined parameter $\beta $ that has the dimensions of an
action (see Eq. (\ref{40})), and rewrite $\widetilde{Q}$ in the general form%
\begin{equation}
\widetilde{Q}(x,z,t)=q_{+}(z_{+},t)q_{-}(z_{-},t)\chi (z_{+},z_{-},t).
\label{56}
\end{equation}%
Note from Eq. (\ref{40}) that $\widetilde{Q}^{\ast }(x,z,t)=\widetilde{Q}%
(x,-z,t)$, so that Eq. (\ref{56}) leads to%
\begin{equation}
q_{+}(z_{-},t)q_{-}(z_{+},t)\chi (z_{-},z_{+},t)=q_{+}^{\ast
}(z_{+},t)q_{-}^{\ast }(z_{-},t)\chi ^{\ast }(z_{+},z_{-},t),  \label{58a}
\end{equation}%
whence%
\begin{equation}
q_{+}(z_{-},t)=q_{-}^{\ast }(z_{-},t),\quad q_{-}(z_{+},t)=q_{+}^{\ast
}(z_{+},t),\quad \chi ^{\ast }(z_{+},z_{-},t)=\chi (z_{-},z_{+},t).
\label{58b}
\end{equation}%
Consequently $\widetilde{Q}(x,z,t)$ can be rewritten as%
\begin{equation}
\widetilde{Q}(x,z,t)=q_{+}(z_{+},t)q_{+}^{\ast }(z_{-},t)\chi
(z_{+},z_{-},t)=q(z_{+},t)q^{\ast }(z_{-},t)\chi (z_{+},z_{-},t),  \label{59}
\end{equation}%
with%
\begin{equation}
q(z_{\pm },t)\equiv q_{+}(z_{\pm },t)=q_{-}^{\ast }(z_{\pm },t).  \label{qu}
\end{equation}%
\thinspace Further, from Eqs. (\ref{59}) and (\ref{42a}) we may write 
\begin{equation}
\rho (x,t)=\widetilde{Q}(x,0,t)=q^{\ast }(x,t)q(x,t)\chi _{0}(x,t),\quad
\label{58e}
\end{equation}%
with $\chi _{0}(x,t)=$ $\left. \chi \right\vert _{z=0}$ a real function that
can be taken as a constant without loss of generality, absorbing its
possible time and space dependence into the functions $q(x,t),$ $q^{\ast
}(x,t)$. Thus 
\begin{equation}
\chi _{0}(x,t)=1,\quad \rho (x,t)=q^{\ast }(x,t)q(x,t).  \label{rho}
\end{equation}

We now return to Eq. (\ref{32}). Its Fourier transform is\footnote{%
In what follows we will assume that all surface terms vanish at infinity.} 
\begin{equation}
\frac{\partial \widetilde{Q}}{\partial t}-i\frac{1}{m}\frac{\partial ^{2}%
\widetilde{Q}}{\partial x\partial z}-izf(x)\widetilde{Q}-\frac{\tau }{m}%
f^{\prime }z\frac{\partial \widetilde{Q}}{\partial z}=-ie^{2}z(\widetilde{%
\hat{D}Q}).  \label{44}
\end{equation}%
To translate this into a description in configuration space that preserves
the information coming from momentum space, we expand Eq. (\ref{44}) into a
power series around $z=0$ and separate the coefficients of $z^{k}$ ($%
k=0,1,2,\ldots $). In this form we obtain an infinite hierarchy of equations
containing moments of $p$ of increasing order. For the first two equations
(coefficients of $z^{0}$ and $z$) we get 
\begin{subequations}
\begin{equation}
\frac{\partial \rho }{\partial t}+\frac{1}{m}\frac{\partial }{\partial x}%
\left( \left\langle p\right\rangle _{x}\rho \right) =0,  \label{46a}
\end{equation}%
\begin{equation}
\frac{\partial }{\partial t}\left( \left\langle p\right\rangle _{x}\rho
\right) +\frac{1}{m}\frac{\partial }{\partial x}(\left\langle
p^{2}\right\rangle _{x}\rho )-f\rho =\frac{\tau }{m}f^{\prime }\left\langle
p\right\rangle _{x}\rho -e^{2}\left. (\widetilde{\hat{D}Q})\,\right\vert
_{z=0}.  \label{46b}
\end{equation}%
The subsequent equations (corresponding to higher powers of $z$) are linked
to the above couple by the same elements $\rho ,\left\langle p\right\rangle
_{x},\left\langle p^{2}\right\rangle _{x}$ and higher-order moments $%
\left\langle p^{n}\right\rangle _{x},$ in addition of course to
contributions deriving from the term $z(\widetilde{\hat{D}Q}).$ Equation (%
\ref{46a}) is the continuity equation (for the transfer of matter) in
configuration space, with the local (partially averaged) flow velocity given%
{\Large \ }by $v(x)=\left\langle \dot{x}\right\rangle _{x}=\tfrac{1}{m}%
\left\langle p\right\rangle _{x}.$ Equation (\ref{46b}) describes the
transfer of momentum density and contains the second moment $\left\langle
p^{2}\right\rangle _{x},$ whose value reappears in the third equation, the
one that describes the transfer of kinetic energy density, and so on. This
coupling between successive equations creates a highly difficult
mathematical problem. However, in the case of interest here, a decoupling of
the first two equations from the rest takes place in a certain
approximation, as shown in the next section. Therefore only the first two%
{\LARGE \ }local moments of $p$ intervene in the resulting (approximate)
description, which means that we may concentrate on the behavior of $%
\widetilde{Q}$ for small values of $z$ (see Eqs. (\ref{42a}) and (\ref{42b}%
)).

Resorting to Eqs. (\ref{42b}) and (\ref{56}) we obtain 
\end{subequations}
\begin{equation}
\left\langle p\right\rangle =-i\left. \left( \frac{1}{\widetilde{Q}}\partial
_{z}\widetilde{Q}\right) \right\vert _{z=0}=-i\beta \left[ \partial _{x}\ln
q(x,t)-\partial _{x}\ln q^{\ast }(x,t)\right] -i\left. \left( \partial
_{z}\ln \chi \right) \right\vert _{z=0}  \label{pp}
\end{equation}%
and for the fluctuations of $p,$ given by $\sigma _{p}^{2}(x)=\left\langle
p^{2}\right\rangle _{x}-\left\langle p\right\rangle _{x}^{2},$ we have 
\begin{equation}
\sigma _{p}^{2}(x)=-\left. \left( \frac{\partial ^{2}}{\partial z^{2}}\ln 
\widetilde{Q}\right) \right\vert _{z=0}=-\beta ^{2}\partial _{x}^{2}\ln \rho
(x,t)+\left. \left[ \left( \beta ^{2}\partial _{x}^{2}-\partial
_{z}^{2}\right) \ln \chi \right] \right\vert _{z=0}.  \label{60aa}
\end{equation}%
An expansion of $\chi (x,z,t)$ as a power series of $z$%
\begin{equation}
\chi (x,z,t)=\tsum_{s=0}^{\infty }z^{s}\chi _{s}(x,t)  \label{xi}
\end{equation}%
allows us to rewrite Eqs. (\ref{pp}) and (\ref{60aa}) as%
\begin{eqnarray}
\left\langle p\right\rangle &=&-i\beta \left[ \partial _{x}\ln
q(x,t)-\partial _{x}\ln q^{\ast }(x,t)\right] -i\chi _{1},  \label{psigmap}
\\
\sigma _{p}^{2}(x) &=&-\beta ^{2}\partial _{x}^{2}\ln q^{\ast
}(x,t)q(x,t)-2\chi _{2}+\chi _{1}^{2}.  \label{psigma2}
\end{eqnarray}

\section{Radiationless approximation in the\newline
time-asymptotic limit}

Equation (\ref{44}), or for that matter, the complete hierarchy of equations
in configuration space, provides in principle an exact statistical
description of the evolution of the ensemble of particles. It is clear that
initially, when particle and field start to interact, there is an
irreversible process of energy and momentum exchange during which the field
has a randomizing and dissipative effect on the particle dynamics, due to
the diffusion and radiation reaction terms (proportional to $e^{2}$ and $%
\tau \sim e^{2},$ respectively) in Eq. (\ref{44}). However, we are here
specifically interested in the time-asymptotic limit, when the combined
action of these two terms has led to a stationary regime in which this
exchange is no more irreversible. In this situation a balance should be
attained between the mean power radiated and that absorbed by the particle
from the field; this balance condition, discussed in detail in section \ref%
{balance}, means that the equality $\frac{\tau }{m}\left\langle f^{\prime
}\left\langle p\right\rangle _{x}\rho \right\rangle =e^{2}\left\langle
\left. (\widetilde{\hat{D}Q})\,\right\vert _{z=0}\right\rangle $ should hold
(see Eq. (\ref{122}) below). One may then consider that the remaining effect
of the two terms, both proportional to $e^{2}$ and thus of order $\alpha $ $%
=e^{2}/\hbar c$,$^{\text{\cite{dice}}}$ is reduced to a small (stochastic)
radiative correction. By neglecting the terms on the right-hand side of Eq.~(%
\ref{46b}) one obtains 
\begin{subequations}
\label{47}
\begin{eqnarray}
\frac{\partial \rho }{\partial t}+\frac{1}{m}\frac{\partial }{\partial x}%
\left( \left\langle p\right\rangle _{x}\rho \right) &=&0,  \label{47a} \\
\frac{\partial }{\partial t}\left( \left\langle p\right\rangle _{x}\rho
\right) +\frac{1}{m}\frac{\partial }{\partial x}(\left\langle
p^{2}\right\rangle _{x}\rho )-f\rho &=&0.  \label{47b}
\end{eqnarray}

As follows from Eqs. (\ref{psigmap}) and (\ref{psigma2}), the first of these
equations depends on $\chi _{1},$ the second one on $\chi _{1}$ and $\chi
_{2},$ the third one on $\chi _{1}$, $\chi _{2}$ and $\chi _{3},$ and so on,
so the functions $\chi _{s}$ ($s=1,2,...)$ --- the non-factorizable part of $%
\widetilde{Q}$ (see Eq. (\ref{56}))--- successively couple the equations of
the hierarchy. Before proceeding to the decoupling announced above, it is
convenient to distinguish for a moment between the canonical and the
mechanical momentum of the particle. Note that $p$ in Eqs. (\ref{psigmap})
and (\ref{psigma2}) refers to the mechanical momentum, $p_{\text{m}}=p=P-%
\tfrac{e}{mc}A$ $,$ where $A$ stands for the $x-$component of the
electromagnetic potential $\mathbf{A}$ (including radiation reaction),
whence 
\end{subequations}
\begin{eqnarray}
\left\langle p\right\rangle _{x} &=&\left\langle P\right\rangle _{x}-\tfrac{e%
}{mc}\left\langle A\right\rangle _{x},  \label{disp} \\
\sigma _{p}^{2}(x) &=&\sigma _{P}^{2}(x)-\tfrac{e^{2}}{m^{2}c^{2}}\Sigma
(A,p)  \notag
\end{eqnarray}%
where $\Sigma (A,p)=\sigma _{A}^{2}(x)+\left( 2mc/e\right) \Gamma _{Ap}$,
and $\Gamma _{AB}=\left\langle AB\right\rangle _{x}-\left\langle
A\right\rangle _{x}\left\langle B\right\rangle _{x}.$ Now, resorting to Eqs.
(\ref{psigmap}) and (\ref{psigma2}) and using (\ref{58e}), one is led to 
\begin{subequations}
\begin{eqnarray}
\left\langle P\right\rangle _{x}-\tfrac{e}{mc}\left\langle A\right\rangle
_{x} &=&-i\beta \partial _{x}\ln \frac{q(x,t)}{q^{\ast }(x,t)}-i\chi _{1},
\label{pAbis} \\
\sigma _{P}^{2}(x)-\tfrac{e^{2}}{m^{2}c^{2}}\Sigma (A,p) &=&-\beta
^{2}\partial _{x}^{2}\ln q^{\ast }(x,t)q(x,t)-2\chi _{2}+\chi _{1}^{2}.
\label{sigmabis}
\end{eqnarray}%
\ At this point we \textit{assume} (this is our Ansatz) that the (local)
moments of $P$ are determined by the functions $q(x,t),$ $q^{\ast }(x,t)$,
so that 
\end{subequations}
\begin{eqnarray}
\left\langle P\right\rangle _{x} &=&-i\beta \left[ \partial _{x}\ln
q(x,t)-\partial _{x}\ln q^{\ast }(x,t)\right] ,  \label{p media} \\
\sigma _{P}^{2}(x) &=&-\beta ^{2}\partial _{x}^{2}\ln q^{\ast }(x,t)q(x,t).
\label{sigma}
\end{eqnarray}%
This means, in particular, that the information regarding the phase of $%
q(x,t)$ is transferred to the local average of the canonical momentum and
the information regarding the magnitude of $q(x,t)$ is transferred to the
dispersion of the canonical momentum. The above assumption implies that the
terms containing $\chi _{1}$ and $\chi _{2}$ in Eqs. (\ref{pAbis}) and (\ref%
{sigmabis}) are the ones that bear information regarding the electromagnetic
potential $A$, so that once energy balance has been reached the following
equalities should hold: 
\begin{equation}
-i\left. \left( \partial _{z}\ln \chi \right) \right\vert _{z=0}=-i\chi
_{1}=-\tfrac{e}{mc}\left\langle A\right\rangle _{x},\quad 2\chi _{2}=\tfrac{%
2e}{mc}\Gamma _{Ap}-\tfrac{e^{2}}{m^{2}c^{2}}\left\langle A^{2}\right\rangle
_{x}.  \label{third cond}
\end{equation}%
Similarly, the remaining terms $\chi _{s}(x,t)$ with $s\geq 3$ in Eq. (\ref%
{xi}) should represent $A$-dependent contributions to the higher moments of $%
p$. This is equivalent to assuming that the entire function $\chi $
contributes only radiative corrections to the resulting equations, hence in
the radiationless approximation we should take $\chi \simeq \chi _{0}=1$.
This is well verified a posteriori, since $\chi $ leads to the Lamb shift of
the energy levels, a very small correction indeed.$^{\text{\cite{CePe88}}}$
Thus in what follows we take the radiationless approximation $P\rightarrow p$
and $\chi =1,$ whence 
\begin{equation}
\left\langle p\right\rangle _{x}=mv(x)=-i\beta \left[ \partial _{x}\ln
q(x,t)-\partial _{x}\ln q^{\ast }(x,t)\right] ,  \label{131}
\end{equation}%
and, using Eq. (\ref{rho}),%
\begin{equation}
\sigma _{p}^{2}(x)=\left\langle p^{2}\right\rangle _{x}-\left\langle
p\right\rangle _{x}^{2}=-\beta ^{2}\partial _{x}^{2}\ln q^{\ast
}(x,t)q(x,t)=-\beta ^{2}\partial _{x}^{2}\ln \rho ,  \label{60c}
\end{equation}%
or $\left\langle p^{2}\right\rangle _{x}=\left\langle p\right\rangle
_{x}^{2}-\beta ^{2}\partial _{x}^{2}\ln \rho .$ This means that $%
\left\langle p^{2}\right\rangle _{x}$ has become a function of $\left\langle
p\right\rangle _{x}$ and $\rho $ only, so that the first two equations of
the hierarchy suffice to determine the two unknown functions $\rho (x,t)$
and $\left\langle p\right\rangle _{x}.$ Equations (\ref{47a}), (\ref{47b})
have decoupled from the rest and become an independent system of equations
for the evolution of the ensemble.

In the following section we tackle the problem of solving this system of
equations. This is a complicated nonlinear system of equations, so the
search for a simpler procedure is an important task. It turns out to be
important indeed, since the end result is a couple of linear equations, as
shown in the next section.

\section{The quantum description}

In order to determine $q(x,t)$ we resort to Eq. (\ref{44}), which in the
current (radiationless) approximation reduces to%
\begin{equation}
\frac{1}{\widetilde{Q}}\partial _{t}\widetilde{Q}-\frac{i}{m\widetilde{Q}}%
\partial _{x}\partial _{z}\widetilde{Q}=izf(x).\text{ }  \label{A6}
\end{equation}%
In this equation we should take $\chi =1$ according to our previous
observations; however we will write still $\widetilde{Q}=q_{+}q_{-}\chi $
(according to Eq. (\ref{56})), leaving explicit for a moment the $\chi $
function, because it is instructive to find the role it plays in a crucial
place. We thus obtain%
\begin{equation*}
\frac{1}{\chi }\partial _{t}\chi +\frac{1}{q_{+}q_{-}}\partial
_{t}(q_{+}q_{-})-\frac{i\beta }{m}\left[ \frac{1}{q_{+}\chi }\partial
_{+}^{2}(q_{+}\chi )-\frac{1}{q_{-}\chi }\partial _{-}^{2}(q_{-}\chi )\right]
=
\end{equation*}%
\begin{equation}
=\frac{i}{2\beta }(z_{+}-z_{-})f[\tfrac{1}{2}(z_{+}+z_{-})].\text{ }
\label{A12}
\end{equation}%
For sufficiently small values of $z=\left( z_{+}-z_{-}\right) /2\beta $
(corresponding to the first moments of $p$), the mean-value theorem can be
applied to the term on the second line of Eq. (\ref{A12}) 
\begin{equation}
\int_{z_{-}}^{z_{+}}f(u)du=(z_{+}-z_{-})f\left[ \tfrac{1}{2}\left(
z_{+}+z_{-}\right) +\tfrac{1}{2}\epsilon (z_{+}-z_{-})\right] ,\text{ }
\label{A14}
\end{equation}%
where $\epsilon $ is a number between $-1$ and $+1$. Assuming $f(z)$ to be
well behaved in the small interval $[z_{-},z_{+}],$ we have to lowest order
in $z$ 
\begin{equation}
(z_{+}-z_{-})f[\tfrac{1}{2}(z_{+}+z_{-})]=\int_{z_{-}}^{z_{+}}f(u)du=-\left[
V(z_{+})-V(z_{-})\right] ,\text{ }  \label{A16}
\end{equation}%
where $V(z)$ is the potential associated with the force $f(z).$ This result
holds for any (well-behaved) potential and not just for quadratic functions,
for which it is immediate. Equation (\ref{A12}) now takes on the form%
\begin{equation*}
\frac{1}{\chi }\partial _{t}\chi +\frac{1}{q_{+}q_{-}}\partial
_{t}(q_{+}q_{-})-\frac{i\beta }{m}\left[ \frac{1}{q_{+}\chi }\partial
_{+}^{2}(q_{+}\chi )-\frac{1}{q_{-}\chi }\partial _{-}^{2}(q_{-}\chi )\right]
=
\end{equation*}%
\begin{equation}
=-\frac{i}{2\beta }\left[ V(z_{+})-V(z_{-})\right] .  \label{A18}
\end{equation}%
In terms of the couple of functions 
\begin{equation}
\Psi _{+}(z_{+},z_{-},t)=q_{+}(z_{+},t)\chi (z_{+},z_{-},t),\quad \Psi
_{-}(z_{+},z_{-},t)=q_{-}(z_{-},t)\chi (z_{+},z_{-},t),  \label{A23}
\end{equation}%
equation (\ref{56})\ reads%
\begin{equation}
\widetilde{Q}(x,z,t)=\frac{\Psi _{+}\Psi _{-}}{\chi },  \label{A28}
\end{equation}%
and Eq. (\ref{A18}) becomes{\LARGE \ } 
\begin{gather}
\frac{1}{\Psi _{+}}\left[ -\frac{i\beta }{m}\frac{\partial ^{2}\Psi _{+}}{%
\partial z_{+}^{2}}+\frac{i}{2\beta }V(z_{+})\Psi _{+}+\frac{\partial \Psi
_{+}}{\partial t}\right] +  \notag \\
+\frac{1}{\Psi _{-}}\left[ \frac{i\beta }{m}\frac{\partial ^{2}\Psi _{-}}{%
\partial z_{-}^{2}}-\frac{i}{2\beta }V(z_{-})\Psi _{-}+\frac{\partial \Psi
_{-}}{\partial t}\right] =\frac{1}{\chi }\frac{\partial \chi }{\partial t}.
\label{A32}
\end{gather}%
In this equation, the functions $\Psi _{+}$ and $\Psi _{-}$ are coupled
through $\chi (z_{+},z_{-},t)$ and will remain so as long as the latter
plays any significant role. In the radiationless approximation the
right-hand side reduces to zero and Eq. (\ref{A32}) becomes separable. This
is the crucial place referred to above: it is only in the radiationless
approximation that the functions $\Psi _{+}$ and $\Psi _{-}$ obey each one a
separate equation. Knowing this we take the limit $z\rightarrow 0$ (so that
both $z_{+}$ and $z_{-}$ reduce to $x$), $\chi =1$ and using Eq. (\ref{58b})
put 
\begin{equation}
\left. \Psi _{+}\right\vert _{z=0}=q_{+}(x,t)\equiv e^{iCt}\psi (x,t),\quad
\left. \Psi _{-}\right\vert _{z=0}=q_{+}^{\ast }(x,t)=e^{-iCt}\psi ^{\ast
}(x,t),  \label{psi}
\end{equation}%
with $C$ the constant of separation. Eq. (\ref{A32}) separates and we
finally get 
\begin{equation}
-2\frac{\beta ^{2}}{m}\frac{\partial ^{2}\psi }{\partial x^{2}}+V(x)\psi
=2i\beta \frac{\partial \psi }{\partial t}  \label{A44}
\end{equation}%
and its complex conjugate. The particle density $\rho $ is now given by%
{\LARGE \ }%
\begin{equation}
\rho (x,t)=\psi ^{\ast }(x,t)\psi (x,t),  \label{A35}
\end{equation}%
as follows from Eqs. (\ref{rho}) and (\ref{psi}). Except for the (constant)
factor $\beta $ to be determined (see the discussion in section \ref{balance}%
), we recognize in (\ref{A44}) the Schr\"{o}dinger equation for the (wave)
function $\psi (x,t)$.

According to the above discussion, Eqs. (\ref{A44}) and (\ref{A35}) provide
a valid description in the particle's configuration space only once the
combined irreversible effect of the radiative terms has taken place, energy
balance has been reached and the radiationless approximation can be taken.
Under these circumstances all (reversible) time evolution is confined to the
time dependence of the wave functions$\ \psi $ and $\psi ^{\ast }.$ In a
more exact description (the \textsc{qed} description, in particular) it will
be necessary to make recourse to the remaining equations of the hierarchy to
extract the additional information that has been left out. In particular and
as was already stated, it can be demonstrated by a perturbative calculation
that the terms that depend on $\chi _{1}$ and $\chi _{2}$ give rise to the
Lamb shift of the atomic levels plus a nonrelativistic mass correction.$^{%
\text{\cite{PeVaCe09},\cite{CePe88}}}$

One can consider the above results as a determination of the assumptions,
conditions and approximations required to arrive at quantum mechanics from
an initially complete phase-space description. From such a perspective this
work serves to disclose some of the hidden features behind the quantum
phenomenon. Whether in all cases the system under study reaches a regime in
which the above conditions hold, remains to be studied. Whenever they do
reach it, however, these systems are of utmost importance: they are just the
quantum systems.

\subsection{The quantum potential\label{QP}}

Resorting to Eq. (\ref{60c}) we can rewrite the couple of equations (\ref%
{47a}) and (\ref{47b}), with $v=\tfrac{1}{m}\left\langle p\right\rangle _{x}$
as{\LARGE \ } 
\begin{subequations}
\label{70vv}
\begin{gather}
\frac{\partial \rho }{\partial t}+\frac{\partial }{\partial x}\left( v\rho
\right) =0,  \label{70a} \\
m\frac{\partial }{\partial t}\left( v\rho \right) +m\frac{\partial }{%
\partial x}\left( v^{2}\rho \right) -\frac{\beta ^{2}}{m}\frac{\partial }{%
\partial x}\left( \rho \frac{\partial ^{2}}{\partial x^{2}}\ln \rho \right)
-f\rho =0.  \label{70b}
\end{gather}%
According to the above discussions, when the radiationless approximation is
taken and the original (statistical) description is reduced to configuration
space,\ it is not a Fokker-Planck-type equation what controls the evolution
of the system, but the couple of equations (\ref{70vv}) instead, or,
equivalently, the Schr\"{o}dinger equation.\footnote{%
An important remark by Wallstrom (1994) about this equivalence is briefly
discussed in section \ref{balance}.} In these equations the only (explicit)
remaining connection to momentum space is the term containing $\ln \rho .$
As Eq. (\ref{60c}) shows, this contribution comes from the local value of
the fluctuations of the momentum: a term that is unusual in classical
mechanics, but essential in quantum physics. If we write the momentum $p$\
of one specific particle of the ensemble at a given point $x$\ in the form 
\end{subequations}
\begin{subequations}
\begin{equation}
\left. p\right\vert _{x}=mv(x)+\delta (x),  \label{76a}
\end{equation}%
where $\delta (x)$ stands for the deviation, then 
\begin{equation}
\sigma _{p}^{2}(x)=\left\langle \delta ^{2}(x)\right\rangle _{x}.
\label{76b}
\end{equation}%
The momentum fluctuation averaged over the whole phase space with
probability density $Q(x,p,t)$ is given by 
\end{subequations}
\begin{eqnarray}
\left\langle \sigma _{p}^{2}(x)\right\rangle &=&\int \left\langle \delta
^{2}(x)\right\rangle _{x}\rho (x)dx  \notag \\
&=&-\beta ^{2}\int \rho \frac{\partial ^{2}}{\partial x^{2}}\ln \rho
dx=\beta ^{2}\int \rho \left( \frac{1}{\rho }\frac{\partial \rho }{\partial x%
}\right) ^{2}dx,  \label{77}
\end{eqnarray}%
or%
\begin{equation}
\left\langle \sigma _{p}^{2}(x)\right\rangle =\beta ^{2}\left\langle \left( 
\frac{1}{\rho }\frac{\partial \rho }{\partial x}\right) ^{2}\right\rangle .
\label{78}
\end{equation}%
This formula (when multiplied by $1/2m$) represents a central contribution
to the average kinetic energy of the particle. It is important to recognize
that the centered momentum $\sigma _{p}^{2}(x)$ at a given point $x$ bears
statistical information about the entire (available) space, due to its
dependence on the probability density $\rho (x)$, and thus it introduces a
nonlocal ingredient into the description through Eq. (\ref{70b}).\footnote{%
\label{unote}This contribution appears in the literature under several
guises. It is discussed in Refs. \cite{Wa89}, \cite{Wa94}, where attention
is paid to its nonlocal nature. In stochastic quantum mechanics it is
identified as produced by the stochastic velocity $u=\left( \hbar /2m\right)
\left( \partial _{x}\rho /\rho \right) $ (see, e.g., Refs. \cite{Ne85}-\cite%
{PeCe82}, \cite{dice} and references therein). This same contribution to the
kinetic energy is interpreted by Olavo as coming from a local entropy due to 
\emph{spontaneous} local fluctuations in positions (see Ref. \cite{Ol00}).
In Bohmian quantum mechanics, the causal version of this theory, extensive
use is made of it, giving rise to the quantum potential (e.g. Ref. \cite%
{Ho93}); see Eq. (\ref{94a}) below.
\par
Note that the momentum associated with the velocity $u$ is given by $\beta
\partial _{x}\rho /\rho $, hence it follows from Eq. (\ref{78}) that it is
directly related to $\left\langle \sigma _{p}^{2}(x)\right\rangle ^{1/2}$.}%
{\large \ }

To establish the connection between this term and the quantum potential we
write the solution of Eq. (\ref{A44}) in the familiar form 
\begin{equation}
\psi (x,t)=\sqrt{\rho }e^{iS(x,t)},  \label{92a}
\end{equation}%
whence 
\begin{equation}
v(x)=\frac{1}{m}\left\langle p\right\rangle _{x}=\frac{2\beta }{m}\frac{%
\partial S}{\partial x},  \label{92b}
\end{equation}%
as follows from Eq. (\ref{131}). Equation (\ref{70b}) rewritten with the aid
of Eq. (\ref{92b}) gives after an integration (the integration constant is
absorbed in $S$) 
\begin{equation}
2\beta \frac{\partial S}{\partial t}+\frac{2\beta ^{2}}{m}\left( \frac{%
\partial S}{\partial x}\right) ^{2}-\frac{2\beta ^{2}}{m}\frac{1}{\sqrt{\rho 
}}\frac{\partial ^{2}\sqrt{\rho }}{\partial x^{2}}+V=0,  \label{94a}
\end{equation}%
where $V$ is the potential associated with the external force $f(x).$ Except
for the third term (originating in the term that contains $\ln \rho $ in Eq.
(\ref{70b})) this is the Hamilton-Jacobi equation of classical mechanics,
with $S$ playing the role of the action function. Hence the statistical (and
nonlocal) nature of the description of the ensemble is encapsulated in this
term alone, just the one that contains the information about the
fluctuations impressed by the zpf and marks the deviation of the system from
its classical behaviour. Note also that this equation (together with Eq. (%
\ref{92b}), known as the \textquotedblleft guiding\textquotedblright\
formula) is precisely the starting point in the usual presentations of
Bohmian quantum mechanics.\footnote{%
As follows from Eq. (\ref{76a}), Eq. (\ref{92b}) refers to the local mean
velocity, not to the individual local velocity of a given particle. Hence it
seems necessary to add a stochastic component $\delta v(x)$ to $v(x)$ in
Bohmian mechanics to get the true individual (stochastic) velocities. See
e.g. Refs. \cite{Ho93}, \cite{DuGoZa92}.} The `extra'\ term is known in such
context as the \emph{quantum potential} (e.g. Ref. \cite{Ho93}, Ch. 3),
although it is of \emph{kinetic} origin. As already stated, this is the main
source of quantum nonlocality (due to its dependence on the spatial
distribution), even for a single particle, arising at the level of the
reduced configuration-space description.\ As is well known, it is also what
makes Bohm's theory nonlocal.

Notice that for any solution of the Schr\"{o}dinger equation (with the
single exception of the free particle described by a plane wave) the quantum
potential is different from zero. If the system is composed of two or more
particles described by a nonfactorizable probability density, there are
additional (interparticle) nonlocal\ contributions to the quantum potential,
also having no classical analog. For instance, for a two-particle density $%
\rho (x_{1},x_{2})=$ $\rho _{1}(x_{1})\rho _{2}(x_{2})\rho
_{12}(x_{1},x_{2}) $ the quantum potential for particle 1 has the form%
\begin{equation}
Q_{1}(1,2)=-\frac{2\beta ^{2}}{m}\left[ \frac{1}{\sqrt{\rho _{1}}}\frac{%
\partial ^{2}\sqrt{\rho _{1}}}{\partial x_{1}^{2}}+\frac{1}{\sqrt{\rho _{12}}%
}\frac{\partial ^{2}\sqrt{\rho _{12}}}{\partial x_{1}^{2}}+\frac{1}{2}\left( 
\frac{\partial }{\partial x_{1}}\ln \rho _{1}\right) \left( \frac{\partial }{%
\partial x_{1}}\ln \rho _{12}\right) \right] .  \label{96}
\end{equation}%
The first term corresponds to the usual single-particle quantum potential;
the remaining two terms, however, are entangled contributions which produce
the nonlocal effects characteristic of Bell's correlations.

\subsection{Some comments on quantum nonlocality}

We have reached a fundamental conclusion, namely that any quantum system
manifests nonlocality, the only possible exception being the free particle
described by a plane wave. The existence of quantum nonlocalities is a
well-known fact. A treatment of them is to be found in the literature
related to the causal version of quantum mechanics proposed by Bohm, a
theory in which the quantum potential plays a fundamental role. However, the
most extensive line of research on quantum nonlocalities during the last
decades is related to the Bell inequalities, a fact that unfortunatedly has
led in some circles to the misleading idea that nonlocality is a property
exclusive of multipartite quantum systems. It should be stressed that,
independently of interpretation, the Schr\"{o}dinger equation contains the
quantum potential --- even if not in an explicit form --- and hence the
associated quantum nonlocalities. What we have demonstrated here is that the
so-called potential --- really a kinetic contribution --- results from the
reduction of the statistical description in the phase space of the particle
to its configuration space. In this space the nonlocality is present,
whereas the original description in the full phase space is as local as any
statistical description can be.

As for multiparticle systems, as has just been seen, extra nonlocalities
arise due to the correlations among variables; these are related to the Bell
nonlocalities. The mechanism that gives rise to the entanglement between the
components of a bipartite system within the present theory is discussed in
detail in Refs. \cite{Va10}, \cite{VaPeCe11}{\footnotesize .}

The cause or origin of both the quantum fluctuations and quantum nonlocality
has been a long standing problem. \textit{What} is fluctuating and \textit{%
why} remains undefined in quantum mechanics. Proposals on the origin of the
quantum potential --- the{\LARGE \ }recognized source of nonlocality --- can
be counted by the dozen (see e.g. Refs. \cite{Ne85}-\cite{Ho93} and \cite%
{Carr10}), the mechanism put forward being frequently alien to quantum
theory. Here we discover a simple and unifying answer: what fluctuates is
the momentum of the particle due to the direct action of the fluctuating
electromagnetic vacuum, and these fluctuations give rise to the quantum
potential when transferred to configuration space.

It could still be argued that local theories cannot lead to nonlocal
results. Of course in the realm of pure dynamics this is true. But the
statistical account of an originally local description can have nonlocal
features. Take as an example the Brownian motion of particles immersed in a
fluid: the (partially averaged) local velocities and accelerations required
for the statistical description in configuration space are determined by the
full phase-space density. Thus, local laws can lead to nonlocalities
associated with the (reduced) statistical description. This is the case in
the present theory, that deals with particles immersed in a local field,
solution of Maxwell's equations.

From this perspective, quantum nonlocality does not refer to an ontological
property, i.e., an ingredient of Nature, but to a property of the (quantum)
description. It is a real nonlocality in the restricted --- but operational
--- sense that it can be \textquotedblleft observed\textquotedblright , if
the description of the observed phenomenon is made within quantum mechanics.
Then the predictions of the theory correspond to what we observe. The point
is how we describe (and interpret) what we observe. In a full phase-space
theory we would employ a local language; in the reduced, quantum mechanical
description, we require a nonlocal one.

Of course ours is not the first and only theory from which it follows that
the violation of Bell's inequalities does not rule out local realism. For
example, since long ago it has been argued by a number of authors that the
violation of a Bell inequality is due to the failure not of locality, but of
the assumption of joint measurability; see in particular Refs. \cite{SuZa81}
and \cite{Br93}. More recent excellent examples are advanced by Khrennikov
(see e.g. Refs. \cite{Kh02}-\cite{Kh09}) and T. M. Nieuwenhuizen,$^{\text{%
\cite{Nieu11}}}$ as well as by 't Hooft in a somewhat different direction.$^{%
\text{\cite{Hooft11}}}$

\subsection{Energy balance and the value of $\protect\beta \label{balance}$}

Let us now look into the condition required to attain a state of constant
mean energy. From Eq. (\ref{32}) it follows, assuming again that all surface
terms vanish at infinity, that 
\begin{subequations}
\begin{equation}
\frac{1}{2m}\frac{d}{dt}\left\langle p^{2}\right\rangle =\frac{1}{2m}\frac{d%
}{dt}\int p^{2}Qdxdp=\frac{1}{m}\left\langle fp+\frac{\tau }{m}f^{\prime
}p^{2}-\frac{e^{2}}{2}p\hat{D}\right\rangle .  \label{120a}
\end{equation}%
Since $\left\langle fp\right\rangle /m=-d\left\langle V\right\rangle /dt,$
the average energy gained or lost by the quantum-mechanical system through
radiation exchange is given by 
\end{subequations}
\begin{equation}
\frac{d}{dt}\left\langle H\right\rangle =\frac{d}{dt}\left\langle \frac{1}{2m%
}p^{2}+V\right\rangle =\frac{\tau }{m^{2}}\left\langle f^{\prime
}p^{2}\right\rangle -\frac{e^{2}}{2m}\left\langle p\hat{D}\right\rangle ,
\label{122}
\end{equation}%
where $H$ represents the mechanical Hamiltonian of the particle. The first
term on the right-hand side of Eq. (\ref{122}) gives the average power
dissipated by the particle along its orbit due to Larmor radiation. The last
term, on the other hand, represents the mean power absorbed by the
fluctuating particle from the field. In particular, when these terms cancel
one another, namely for those averaged motions such that 
\begin{equation}
\frac{\tau }{m^{2}}\left\langle f^{\prime }p^{2}\right\rangle =\frac{e^{2}}{%
2m}\left\langle p\hat{D}\right\rangle ,  \label{124}
\end{equation}%
the value of $\left\langle H\right\rangle $ becomes a constant and an energy
balance is reached. Note, incidentally, that this equation{\tiny \ }plays
the role of a fluctuation-dissipation relationship for the stochastic
(quantum) system.

In particular, when both the background field and the mechanical system are
in their ground state, there is no net exchange of energy and Eq. (\ref{124}%
) is satisfied. This equation can then be used to determine the value of the
parameter $\beta $, as follows. The calculation of the left-hand side to
lowest order in $e^{2}$ (or $\tau $) can be done by resorting to the
(zero-order) solutions of Eq. (\ref{A44}), which leads to a result that
contains the linear factor $\beta $. The right-hand side, on the other hand,
is a complicated expression containing second and higher moments of the
electric component of the background field. However, to lowest order in $%
e^{2}$ the term on the right-hand side becomes proportional to the spectral
energy density of the zpf, which as Eq. (\ref{34}) shows, is linear in
Planck's constant $\hbar $. The detailed calculation, to be presented in a
paper in preparation,$^{\text{\cite{CePeVa}}}$ then leads directly to the
result 
\begin{equation}
\beta =\tfrac{1}{2}\hbar .  \label{125}
\end{equation}%
This completes the identification of Eq. (\ref{A44}) with the Schr\"{o}%
dinger equation.\footnote{%
An informal argument that justifies the result $\beta =\hbar /2$ can be seen
in \cite{PeVaCeFr11}.} Further to indicating the point of entry of Planck's
constant $\hbar $ into the quantum description, the derivation leading to
Eq. (\ref{125}) exhibits the unique role that the zpf, with the specific
spectrum given by Eq. (\ref{34}), plays in achieving energy balance in the
quantum case.

Some time ago Wallstrom$^{\text{\cite{Wa89}, \cite{Wa94}}}$ criticized
theories in which the Schr\"{o}dinger equation is derived from the pair of
equations (\ref{70vv}), arguing that the theory described by this couple of
equations allows in general for more solutions than those afforded by the
corresponding Schr\"{o}dinger equation. However this critique does not apply
to the present case because in addition to Eqs. (\ref{70vv}), the system
must comply with Eq. (\ref{124}), the energy balance condition that can be
satisfied in general only by a discrete set of solutions. Quantization has
therefore a deep physical root within the present approach. Although the
fundamental equations in various phenomenological accounts of quantum
mechanics --- such as Bohmian mechanics,$^{\text{\cite{Ho93}, \cite{BoHi93}}%
} $ Nelson's theory$^{\text{\cite{Ne85}}}$ or stochastic quantum mechanics$^{%
\text{\cite{Pe69}}}$ --- match with the present ones, it is the crucial
condition provided by Eq. (\ref{124}) what makes the difference. For an
additional answer to Wallstrom's criticism see Ref. \cite{Smo06}.

\subsection{The extremum value of the energy}

It is appropriate here to recollect a suggestive derivation related to Eq. (%
\ref{60c}).$^{\text{\cite{PeCe77}}}$ Let us consider a stationary state with 
$v=0;$ equation (\ref{60c}) gives then 
\begin{equation}
\left\langle p^{2}\right\rangle _{x}=-\frac{\hbar ^{2}}{4}\frac{\partial ^{2}%
}{\partial x^{2}}\ln \rho .  \label{136}
\end{equation}%
The mean energy of the mechanical system is given by the mean value of its
Hamiltonian, thus 
\begin{equation}
\left\langle H\right\rangle =\int \rho (x)\left[ -\frac{\hbar ^{2}}{8m}\frac{%
\partial ^{2}}{\partial x^{2}}\ln \rho +V(x)\right] dx.  \label{140}
\end{equation}%
We demand now this energy to be an extremum under conservation of
probability,{\LARGE \ }$\int \rho (x)dx=1.$ With the probability density
written in the form $\rho =\psi ^{2}$ (which is consistent with the
assumption $v=0$), this variational problem has as solution the
Euler-Lagrange equation$^{\text{\cite{Ha99}}}$%
\begin{equation}
-\frac{\hbar ^{2}}{2m}\frac{\partial ^{2}\psi }{\partial x^{2}}+V\psi
=\left\langle H\right\rangle \psi ,  \label{144}
\end{equation}%
where $\left\langle H\right\rangle $ is the Lagrange multiplier.{\LARGE \ }%
This result emphasizes the remarkable role played by the local dispersion of
the momentum, Eq. (\ref{136}): it guarantees that the stationary (quantized)
distribution of {\tiny \ }particles corresponds to a local extremum
(normally a minimum) of the mean energy of the system. We thus discover that
the stochastic velocity $u$ referred to above (see footnote \ref{unote}) and
well known in the context of stochastic quantum mechanics$^{\text{\cite{Ne85}%
, \cite{Pe69}}}$%
\begin{equation}
u=\frac{\beta }{m}\frac{\partial \ln \rho }{\partial x}=\frac{\beta }{m\rho }%
\frac{\partial \rho }{\partial x}=\frac{\hbar }{2m}\left( \frac{1}{\psi
^{\ast }}\frac{\partial \psi ^{\ast }}{\partial x}+\frac{1}{\psi }\frac{%
\partial \psi }{\partial x}\right) ,  \label{146}
\end{equation}%
is instrumental in leading to the quantum description through its
divergence, as follows from the relation%
\begin{equation}
-\frac{1}{2}\beta \frac{\partial u}{\partial x}=-\frac{\beta ^{2}}{2m}\frac{%
\partial ^{2}\ln \rho }{\partial x^{2}}=\frac{\left\langle
p^{2}\right\rangle _{x}}{2m}.  \label{148}
\end{equation}%
Equation (\ref{146}) shows that the gradient of the probability density is
intimately related to the local description of the stochastic component $%
mu\rho $ of the momentum density. Hence the description of the velocity in
terms of the function $\left\langle p\right\rangle _{x}=mv(x)$ alone does
not settle the question about the quantum motions. Actually, quantum
mechanics deals very smoothly and elegantly with these matters, although not
paying too much attention to the ultimate meaning of the results. In fact,
from Eqs. (\ref{131}) and (\ref{146}) it follows that 
\begin{equation}
-i\hbar \frac{\partial \psi }{\partial x}=m(v-iu)\psi =\hat{p}\psi ,
\label{154}
\end{equation}%
where the last equality has been taken from the usual quantum formalism, $%
\hat{p}=-i\hbar \partial _{x}$. The momentum operator takes thus both
components of the velocity into account, although the velocity $u$ (and even
the \textit{velocity} $v$) remains concealed in the standard formalism.

\section{Why the Wigner distribution is not a true probability density \label%
{wigner}}

Since our starting point, Eq. (\ref{30a}), involved a full description in
the entire particle-field phase space, it is interesting to explore whether
a description in the particle's phase space can still be recovered from the
reduced description of quantum mechanics. For this purpose we invert Eq. (%
\ref{40}) and combine it with Eqs. (\ref{52}) and (\ref{56}), obtaining 
\begin{align}
Q(x,p,t)& =\frac{1}{2\pi }\int \widetilde{Q}(x,z,t)e^{-ipz}dz=  \notag \\
& =\frac{1}{2\pi }\int q_{+}(x+\beta z,t)q_{-}(x-\beta z,t)\chi (x+\beta
z,x-\beta z,t)\,e^{-ipz}dz.  \label{110}
\end{align}%
Equation (\ref{110}) furnishes indeed a true (Kolmogorovian) probability
density in phase space. On the other hand, the probability density proper of
quantum mechanics in the radiationless approximation is not this $Q(x,p,t)$,
as we have seen, but its approximate form $W(x,p,t)$ that ensues from
putting $\widetilde{Q}(x,z,t)=q(z_{+},t)q^{\ast }(z_{-},t)=\psi
(z_{+},t)\psi ^{\ast }(z_{-},t)$ as follows from Eq. (\ref{59}) with $\chi
=1\ $and Eq. (\ref{psi}). We thus obtain instead 
\begin{equation}
W(x,p,t)=\frac{1}{\pi \hbar }\int \psi ^{\ast }(x+y,t)\psi
(x-y,t)e^{-i2py/\hbar }dy,  \label{112}
\end{equation}%
with $y=\beta z=\hbar z/2.$ This is the well-known Wigner phase-space
function (see e.g. Refs. \cite{Wi32}, \cite{Mo49}). However, it must be
noted that whilst in Eq. (\ref{110}) the integral runs from $-\infty $ to $%
+\infty $, the functions $\psi ^{\ast },$ $\psi $ as given by the Schr\"{o}%
dinger equation can be taken as approximate representatives of $q_{+},q_{-},$
respectively, only for small values of $y$ and a constant $\chi ,$ since
only then can Eq. (\ref{A32}) be separated into the equations for $\Psi
_{+}=q_{+}\chi (z_{+},z_{-})$ and $\Psi _{-}=q_{-}\chi (z_{+},z_{-}).$

One cannot take $W$ to be a true Kolmogorovian probability in general. And
indeed, despite its recognized value, it is not, since as is well known it
can take on negative values in some regions of phase space for almost all
states and systems. The right solution to this long-standing problem is of
course to recognize the intrinsic limitation of $W$ that ensues from its
approximate nature and to try revert to the original $Q(x,p,t).$ This,
however, being the solution of the integro-differential equation (\ref{32}),
is a much more complicated function, which evolves with time as the system
approaches the state characterized by quantum mechanics. There remains here
an interesting problem to be explored.

\section{Concluding remarks and discussion}

A first conclusion is that the quantum mechanical description provided by
the Schr\"{o}dinger equation emerges naturally from a Fokker-Planck-type
equation (of infinite order) in phase space, namely Eq. (\ref{32}), under
several simplifying assumptions that can be made only when the entire system
has reached energy equilibrium. The original equation remains outside the
limits of quantum mechanics, the latter emerging as a partially averaged,
asymptotic radiationless theory. As a consequence, even though the numerical
correctness of quantum mechanics is out of question for present-day use, its
physical transparency falls far from the mark. The ultimate reason for this
singular situation resides in the fact that the cause of the quantum
behavior --- identified here with the stochastic zero-point radiation field
in interaction with matter --- remains hidden in the background once the
complete description in phase space is reduced to the particle's
configuration space and the radiationless approximation is taken. This leads
to the well-known abstruse narrative of the quantum world, whose statistical
origin remains hidden and in which apparently noncausal and nonlocal effects
take place. According to the above discussion, the so-called quantum
nonlocality arises as soon as the effects of the fluctuations in momentum
space (due to the zpf) are transferred to configuration space. In the full
phase-space description everything is local; the nonlocality appears only in
the reduced description. Indeed, the Schr\"{o}dinger equation is shown to be
intrinsically nonlocal due to the term containing $\sqrt{\rho }$ in Eq. (\ref%
{94a}), which is but a manifestation (on the mechanical subsystem) of the
stochastic nature of the field. These observations --- which serve to
disclose also the physical cause behind the dispersive nature of the quantum
systems --- alert us not to interpret nonlocality as inherent to the
physical system, but rather as a distinctive feature of the quantum
description.

The present work serves to confirm that the stochastic process underlying
quantum mechanics is of a different nature (and requires befitting methods
to handle it) from that of classical stochastic motions.$^{\text{\cite{dice},%
\cite{Pe69},\cite{PeCe82},\cite{PeCe75}}}$In particular, it is remarkable
that even though there exists a true (integro-differential)
Fokker-Planck-type equation in phase space, the appropriate (radiationless)
description of quantum mechanics can be made in terms of just a couple of
equations, one corresponding to the much simpler continuity equation, the
other being close to the classical Hamilton-Jacobi equation, but with a
statistical term originating in the fluctuations in momentum space, over
which an average has been performed.\footnote{%
By its form, this equation suggests a kind of fluid. This is true of course
only for the one particle case (a 3-D fluid). However in any case a
considerable stretch of imagination is required to interpret it in terms of
a quantum fluid. Still, the formal analogy was stressed by Madelung$^{\text{%
\cite{Ma26}}}$ from the very inception of quantum mechanics and continues to
be used on occasions. See e.g Refs. \cite{OrMaSu96} and \cite{LoWy99}.}

The transition from Eqs. (\ref{47a}), (\ref{47b}) to Eqs. (\ref{70vv}) is
clearly an irreversible procedure: relevant information is lost by
disconnecting the former from the rest of the hierarchy. Once this has been
done one cannot transit from Eqs. (\ref{70vv}) back to Eq. (\ref{32}) on
purely logical steps. In particular, one cannot reconstruct the true
probability density in phase space from the (approximate) Wigner function.
This explains the origin of the large number of existing phase-space
versions of quantum mechanics, each one carrying their correlative
correspondence rules, which become an addition to usual quantum mechanics
(see e.g. Ref. \cite{ZaFaCu05}).

It is appropriate to bear in mind that although the zpf could appear as a
sort of collection of hidden variables, introduced with the aim of
completing the quantum description, this is not the case here, since the zpf
is not an ingredient simply added on top of the quantum-mechanical formalism
to make it deterministic. Quite the contrary: nothing is here added to
quantum mechanics, but quantum mechanics \emph{emerges} from a more general
theory that embodies the zero-point field. The emerging description is then
naturally indeterministic, since in every case the specific realization of
the field is unknown, the description having a statistical nature right from
the outset.\footnote{%
The term \textit{indeterministic} is used consistently in this paper (as in
our work in general) to refer to a description that ignores the specific
realizations in an ensemble -- without this, of course, having any
implications on the causality or locality of the system (at the ontological
level).} It is important to note that the particle (the electron, say)
remains always a particle, the existing wave (the zpf) remaining literally
in the background. The trajectories belong to subensembles characterized by
the local mean velocity $v(x).$

The above theory is in principle applicable to an arbitrary number of
particles. Yet it is important to realize that the quantum nonlocality
discussed here emerges already in the one-particle case. For a problem of
several particles, instead of Eqs. (\ref{30b}) we would have a system of
(originally local, stochastic) equations coupled through the respective
radiation reaction terms.$^{\text{\cite{LL}}}$ This leads frequently to a
nonseparable phase-space distribution function, and eventually to a solution 
$\psi $ that is nonfactorizable. When this occurs, the probability density
(in configuration space) contains several terms, including those describing
interferences among single-particle states. The corresponding quantum
potential contains, in addition to the various single-particle terms, extra
multiparticle contributions produced by correlations among the dynamic
variables, that give rise to additional nonlocalities, as Eq. (\ref{96})
illustrates. A most striking manifestation of such `extra' nonlocalities due
to correlations among the particles is related to Bell%
\'{}%
s theorem.$^{\text{\cite{Bell87}}}$ An explanation of the origin of such
additional nonlocalities in the composite problem, or rather of the
mechanism that entangles the particles, is particularly important. In Refs. 
\cite{PeVaCe10}, \cite{Va10} and \cite{VaPeCe11} it is shown how the zpf can
give rise to the entanglement of two noninteracting particles embedded in
the common background field. Specifically, it is shown that even when there
is no external interaction potential between the particles, their
interaction through the background field can correlate and entangle them in
such a way that apparent nonlocalities arise\ at the level of the quantum
description.

Our results serve to explain the success of several works within stochastic
electrodynamics.\footnote{%
There is indeed an ample series of such successful works; here we have
listed only some recent ones that bear direct relation with the present
work. An extensive list of references to works prior to 1996 is given in
Ref. \cite{dice}.} From among those of more direct interest we recall the
numerical calculations of Cole et al$^{\text{\cite{CoZo03a}-\cite{CoZo04}}}$
leading to a correct prediction of the ground state orbit for the H-atom,
the description of spontaneous transitions$^{\text{\cite{FrFrMa97}}}$ and
more generally radiative corrections,$^{\text{\cite{CePe88}}}$ and other
treatments$^{\text{\cite{PeVaCe09},\cite{VaPeCe11}}}$ that complete the
perspective offered by the present one.

\begin{acknowledgement}
A.V.H. acknowledges financial support from the Consejo Nacional de Ciencia y
Tecnolog\'{\i}a, M\'{e}xico.
\end{acknowledgement}

\end{document}